\documentclass[prl,10pt,a4paper,twocolumn,showpacs,notitlepage]{revtex4}

\usepackage{amsmath}
\usepackage{amssymb}
\usepackage[dvips]{graphicx}
\usepackage{times}

\begin{document}

\title{\textbf{Formal equivalence between Tsallis and extended
Boltzmann-Gibbs statistics}}
\author{N. G. de Almeida}
\email{norton@pesquisador.cnpq.br}
\affiliation{N\'{u}cleo de Pesquisas em F\'{\i}sica, Universidade Cat\'{o}lica de Goi\'{a}%
s, 74.605-220, Goi\^{a}nia (GO), Brazil.}

\begin{abstract}
A formal correspondence between the $q-$distribution obtained from the
Tsallis entropy and non-maxwellian distributions obtained from the
Boltzmann-Gibbs entropy is afforded.
\end{abstract}

\maketitle

As is well known, generalized nonextensive statistics have increasingly
received attention, and the physical needs for departure from
Boltzmann-Gibbs statistical mechanics and thermodynamics have been
investigated by several authors \cite{examples}. Since generalized
nonextensive statistics are related to a new physics, it is a fundamental
point to ask for the necessity of introducing nonextensivity at all. It is
our purpose here to show a formal correspondence between the $\rho _{q}$
distribution associated with Tsallis entropy and the $\rho $ distribution
obtained rightly from the Boltzmann-Gibbs entropy.

Tsallis entropy is defined by \cite{tsallis88}%
\begin{equation}
S_{q}=kTr\frac{\rho -\rho ^{q}}{q-1},  \label{1}
\end{equation}%
where $\rho $ is the density matrix, $k$ is a positive constant and $q\in
\mathbb{R}
$ is the entropic index. Since the rule
\begin{equation}
S_{q}(A+B)=S_{q}(A)+S_{q}(B)+(1-q)S_{q}(A)S_{q}(B)\geqslant 0
\end{equation}%
is verified for two independent systems $A$ and $B$, the $q$ parameter has
been interpreted as a measure of nonextensivity. When $q=1$ the
Boltzmann-Gibbs entropy,
\begin{equation}
S=-kTr\rho \ln \rho \text{,}  \label{2}
\end{equation}%
is recovered and corresponds to the extensive case. When $q<1$ and $q>1$ the
system is said to possess superextensive and subextensive properties,
respectively. The statistical operator $\rho _{q}$ obtained maximizing $%
S_{q} $ subject to the constraints
\begin{equation}
Tr\rho =1\text{,}
\end{equation}%
and%
\begin{equation}
Tr\rho ^{q}H=\left\langle E\right\rangle _{q}\text{,}
\end{equation}%
is
\begin{equation}
\rho _{q}=\widehat{Z}_{q}^{-1}\left[ 1-(1-q)\beta H\right] ^{\frac{1}{1-q}}%
\text{,}  \label{2a}
\end{equation}%
where $H$ is the Hamiltonian, $\beta =1/kT$ $\ $is the absolute temperature,
$k$ is the Boltzmann constant and%
\begin{equation}
\widehat{Z}_{q}=Tr\left[ 1-(1-q)\beta H\right] ^{\frac{1}{1-q}}
\end{equation}%
is the generalized partition function. The hat above $\widehat{Z}_{q}$
remind us of its operational character in contrast with the partition
function $Z_{q}$. Once there is no ambiguity in the other operators
appearing here, there is no need for using hat. Eq.(\ref{2a}) can be
rewritten in the expanded form%
\begin{equation}
\rho _{q}=\widehat{Z}_{q}^{-1}\exp \left[ -\sum\limits_{n=1}^{\infty }\frac{%
(1-q)}{n}^{n-1}\left( \beta H\right) ^{n}\right] .  \label{2b}
\end{equation}

As usual, we want to maximize the Boltzmann-Gibbs entropy Eq.(\ref{2}),
subjected to certain constraints. If we know nothing about the system,
except that $Tr\rho =1$, then in the energy representation where $%
H\left\vert E\right\rangle =E\left\vert E\right\rangle $ all the states are
equally likely, i.e., $P(E)=\left\langle E\right\vert \rho \left\vert
E\right\rangle =constant$. On the other hand, if we know the Hamiltonian $H$
for the system and the corresponding mean energy $\left\langle
E\right\rangle =Tr\rho H$, thus using Lagrange multipliers it is
straightforward to show that the entropic form Eq.(\ref{2}) leads to the
Maxwell-Boltzmann distribution $P(E)=Z^{-1}\exp (-\beta E)$. If additional
measurements are made on the system such that, besides the mean energy, we
also know the mean square, or more generally, the central moments of order $%
p $, \textit{i.e.}, $\left\langle \left( \Delta E\right) ^{p}\right\rangle
=\left\langle (H-\left\langle E\right\rangle )^{p}\right\rangle $, we can
impose this knowledge as additional constraints. Note that, classically, if
we know all order of moments, we can precisely characterize the distribution
function. Thus, let us consider $\left\langle \left( \Delta E\right)
^{n}\right\rangle $, $n$ integer, as new constraints and redefine the energy
level such that now we have the moments
\begin{equation}
\left\langle \left( \Delta E\right) ^{n}\right\rangle =Tr\rho H^{n}.
\label{2bb}
\end{equation}%
When we vary $\rho $ in Eq.(\ref{2}) and in those for the constraints Eq.(%
\ref{2bb}), multiplying each constraint by the undetermined Lagrange
multipliers $\beta _{0}$,$\beta _{1}$,$\beta _{2}$, ...$\beta _{n}$ and
adding the result, we obtain
\begin{equation}
Tr\left( 1+\sum\limits_{n=0}^{\infty }\beta _{n}H^{n}+\ln \rho \right)
\delta \rho =0\text{.}
\end{equation}%
Since all variations are independent and $\delta \rho $ is arbitrary, it
follows the extended (non-maxwellian) distribution%
\begin{equation}
\ln \rho =-1-\sum\limits_{n=0}^{\infty }\beta _{n}H^{n}\text{,}
\end{equation}%
or, equivalently
\begin{equation}
\rho =\widehat{Z}^{-1}\exp (-\sum\limits_{n=1}^{\infty }\beta _{n}H^{n})%
\text{,}  \label{2c}
\end{equation}%
where
\begin{equation}
\widehat{Z}=Tr\exp (-\sum\limits_{n=1}^{\infty }\beta _{n}H^{n})  \label{2cc}
\end{equation}%
is the partition function. In the energy representation, Eq.(\ref{2c}) and
Eq.(\ref{2cc}) reads, respectively,%
\begin{equation}
P(E)=Z^{-1}\exp (-\sum\limits_{n=1}^{\infty }\beta _{n}E^{n})\text{,}
\label{2ccc}
\end{equation}%
and%
\begin{equation}
Z=\sum_{E}\exp (-\sum\limits_{n=1}^{\infty }\beta _{n}E^{n})\text{.}
\label{3}
\end{equation}%
Eq.(\ref{3}) can be used to obtain all the Lagrange multipliers considering
formally $E^{k}$ $=Y_{k}$ as independent variables:%
\begin{equation}
\beta _{k}=-\frac{\partial \ln Z}{\partial E^{k}}\text{.}  \label{L}
\end{equation}

The formal correspondence between the $q-$distribution Eq.(\ref{2b}) and the
extended distribution Eq.(\ref{2c}) can be found, both leading to the same
result, imposing%
\begin{equation}
\beta _{n}=\frac{(1-q)^{n-1}}{n}\beta ^{n}\text{.}  \label{2d}
\end{equation}%
From Eq.(\ref{2d}) we see that, in general, in order to get $\rho =\rho _{q}$%
, an infinite number of constraints is needed. However, if the energy
dispersion is not significative, being the distribution highly concentrated
around the mean energy, or, equivalently, if all the Lagrange multipliers
except $\beta _{1}$ are very small, the sum in Eq.(\ref{2c}) and Eq.(\ref{2b}%
) can be truncated, and $q$ in Eq.(\ref{2d}) will differ only slightly from
the unity. Actually, three decades ago, to solve discrepancies between the
observed and predicted fluxes of solar neutrino, D.D. Clayton \cite%
{clayton74} proposed that the Boltzmann factor $\exp (-\beta E)$ was
modified to $\exp (-\beta E-\beta _{2}E^{2})$, which can be rewritten as
\begin{equation}
P(E)=Z^{-1}\exp \left[ -\beta E-\delta \left( \beta E\right) ^{2}\right]
\text{,}  \label{c1}
\end{equation}%
where $\delta $ is a small parameter needed to fit experimental results. As
Clayton noted, $\delta \sim 0.01$ is enough to explain the solar neutrino
fluxes. The agreement between the extended Boltzmann-Gibbs statistics and
the result by Clayton can be obtained truncating Eq.(\ref{2ccc}) after the
quadratic term:
\begin{equation}
P(E)=Z^{-1}\exp (-\beta _{1}E-\beta _{2}E^{2})\text{,}  \label{c2}
\end{equation}%
with
\begin{equation}
Z=\sum_{E}\exp (-\beta _{1}E-\beta _{2}E^{2})\text{.}
\end{equation}%
In the same way, the result by Clayton follows from Tsallis statistics
choosing $q=1-2\delta $. In this context, since $q<1$ the system is regarded
as superextensive, and the inadequacy of Boltzmann-Gibbs entropy is claimed
\cite{kaniadakis96}. Also, a possible correspondence between the Clayton
parameter and the anomalous diffusion phenomena related to the nonextensive
entropy Eq.(\ref{1}) was discussed in Ref.\cite{kaniadakis98}. Further
discussion following this approach can be found in \cite{moreTS}, all of
them reinforcing the nonextensivity. However, we see that Clayton proposals
can be derived from the Boltzmann-Gibbs entropy simply considering the mean
square energy as an additional constraint. The formal correspondence between
Tsallis and the extended Boltzmann-Gibbs statistics, in this case, is
obtained from Eq.(\ref{2d}): $\beta _{1}=\beta $ and $\beta _{2}=\frac{(1-q)%
}{2}\beta ^{2}$. Note that in our derivation the Clayton parameter acquires
a natural meaning, stemming from the energy fluctuations of the system,
which we consider as a new constraint in the maximization of the
Boltzmann-Gibbs entropy. Besides, as a general rule, the factor correcting
Boltzmann-Gibbs statistics is negligible whenever the dispersion be highly
concentrated around the mean energy, \textit{i.e.}, $\left( \Delta E\right)
^{2}\cong 0$, which is undoubtedly the most common situation. On the other
hand, corrections to the Boltzmann factor are expected whenever the
fluctuations become dominant. Also note that to recover the Boltzmann factor
from the extended Boltzmann-Gibbs entropy, it is necessary to impose that
all moments $\left( \Delta E\right) ^{n}$ become successively small,
implying that the energy does not deviate very much from the mean value.
This is specially evident if we rewrite Eq.(\ref{2ccc}) as%
\begin{equation}
P(E)=Z^{-1}\exp (-\sum\limits_{n=1}^{\infty }\widetilde{\beta }_{n}\left( E-%
\overline{E}\right) ^{n})\text{,}  \label{r}
\end{equation}%
where $\widetilde{\beta }_{n}$ are the new Lagrange multipliers that
maximize the Boltzmann-Gibbs entropy subject to the constraints $%
\left\langle \left( \Delta E\right) ^{n}\right\rangle =\left\langle (H-%
\overline{E})^{n}\right\rangle $. To relate the constraints in Eq.(\ref{r})
with those in Eq.(\ref{2ccc}) we have to solve the following equality%
\begin{equation}
\sum\limits_{n=1}^{\infty }\widetilde{\beta }_{n}\left( E-\overline{E}%
\right) ^{n})=\sum\limits_{n=1}^{\infty }\sum\limits_{k=1}^{n}\frac{n!}{%
\left( n-k\right) !k!}\widetilde{\beta }_{n}E^{k}\left( -\overline{E}\right)
^{n-k})\text{,}  \label{binomial}
\end{equation}%
and, for the special case of the Clayton proposal discussed above, Eq.(\ref%
{r}) reduces to
\begin{equation}
P(E)=Z^{-1}\exp \left[ -\widetilde{\beta }_{1}(E-\overline{E})-\widetilde{%
\beta }_{2}(E-\overline{E})^{2}\right] \text{,}  \label{c3}
\end{equation}%
which, when compared to Eq.(\ref{c1}), gives the new constraints: $%
\widetilde{\beta }_{2}=\delta \beta ^{2}$ and $\ \ \widetilde{\beta }%
_{1}=\beta +2\delta \beta ^{2}\overline{E}$. On the other hand, Eq.(\ref{2a}%
) and its corresponding distribution function in the energy representation,
is obtained from Tsallis entropy with the constraint given solely by the
mean energy, and the Boltzmann factor is recovered from Tsallis entropy
simply establishing the limit $q\rightarrow 1$ without reference to the
variance or the moments $\left( \Delta E\right) ^{n}$. Also, note that, for
all systems having the same absolute temperature, the Lagrange multiplier
associated with the mean energy is the same, that is $\beta _{1}=\beta $,
while, in general, the others Lagrange multipliers $\beta _{n+1}$ will be
different for each system. This is true for the generalized distribution
stemming from Boltzmann-Gibbs entropy as well as for the generalized
distribution stemming from Tsallis entropy, thus giving rise to free
parameter(s). That this must be, in general, true is particularly
illuminating from the approach presented here. In fact, based on Eq.(\ref{r}%
) we can immediately see that, since the second central moment is related to
the mean square deviation, two or more distributions having the same
temperature, \textit{i.e.}, the same mean energy, can be distinguished by
the energy dispersion. However, if the mean energy and its dispersion is not
enough to distinguish between two or more distributions, the third central
moment, being proportional to the coefficient of skewness, which is a
measure of the asymmetry of the probability distribution, can be used. If
even the third central moment is not enough, we can use the fourth central
moment, which is proportional to the kurtosis, and so on. Therefore, we can
conclude that by maximizing the Boltzmann-Gibbs entropy we will find, in
general, that the values for the Lagrange multipliers in Eq.(\ref{L}),
related to energy fluctuations, will depend on the system under
consideration, turning difficult, if not impossible, to assert \textit{a
priori} their values. This same difficulty applies to the $q$ entropic
index, which is argued to be related to the microscopic dynamics intrinsic
to each system.

In summary, we have shown a formal correspondence between Tsallis and the
extended Boltzmann-Gibbs statistics. This formal correspondence is obtained
assuming that the infinite constraints maximizing Boltzmann-Gibbs entropy,
and leading to an extended distribution function, are related to the $q$
entropic index appearing in Tsallis entropy in such a way that both the
distribution function obtained by maximizing Tsallis entropy and the
distribution function obtained by maximizing Boltzmann-Gibbs entropy become
identical. Also, we have suggested a relationship between the $q$ entropic
index and the higher order moments of the extended Boltzmann-Gibbs entropy.
As a final remark, we stress that the formal equivalence between Tsallis and
the extended Boltzmann-Gibbs statistics demonstrated here rises an
interesting question related to a possible pseudo nonextensive situation, a
subject hereafter deserving more attention.

I would like to thank VPG/UCG and CNPq for partial support.

\end{document}